\documentclass[12pt]{article}
\usepackage{amssymb}

\def\a{\alpha}

\def\b{\beta}

\def\o{\omega}

\newcommand{\xx}{{\bf x}}
\def\q{\,,\qquad}

\def\be{\begin{equation}}
\def\ee{\end{equation}}
\def\bea{\begin{eqnarray}}
\def\eea{\end{eqnarray}}

\newcommand{\D}{{\mathcal{D}}}
\newcommand{\ie}{{\it i.e.,\,\,}}
\newcommand{\dr}{{\rm d}}

\newcommand{\cc}{{\cal C}}

\begin{document}

\begin{flushright}
\vspace{1mm}
 FIAN/TD/04--25\\
\vspace{-1mm}
\end{flushright}\vspace{1cm}

\begin{center}
{\large\bf  Dirac Singleton as a Relativistic Field Beyond Standard Model
}
\vglue 0.6  true cm
\vskip0.8cm
{M.A.~Vasiliev}
\vglue 0.3  true cm

I.E.Tamm Department of Theoretical Physics, Lebedev Physical Institute,\\
Leninsky prospect 53, 119991, Moscow, Russia
\vskip1.3cm
\end{center}

\vspace{1.2cm}

\vspace {1cm}

{\it $\phantom{MMMMMMMMMMMMMMMMMM}$ To the memory of Alexei Starobinsky}


\vspace{0.4 cm}

\vspace{0.4 cm}

\begin{abstract}

A new interpretation of  Dirac singletons \cite{Dirac:1963ta}, \ie
free conformal fields in $d$ dimensions,  as
relativistic fields in a $d+1$-dimensional space-time
with cosmological constant, that  differs
 from the Flato-Fronsdal dipole construction in $AdS_{d+1}$ \cite{Flato:1986uh},
 is proposed. The $d+1$-dimensional field is described at the level of both equations and Lagrangian. It forms an infinite-dimensional representation of the $d+1$-dimensional Lorentz group that relates  fields at
different space-time points. The associated well-known fact is that singleton cannot be localized at a point in ${d+1}$ dimensions,
hence being unobservable via local scattering/radiation phenomena in the
Standard Model  ($d=3$).
 On the other hand, that singleton respects ${d+1}$ dimensional  relativistic symmetries makes it possible to introduce its interactions with gravity and other relativistic  fields in $d+1$ dimensions.
 It is speculated that the presence of singleton in a four-dimensional field theory with non-zero cosmological constant (dark energy) can  be relevant to
  the dark matter phenomenon and baryon asymmetry generation.

\end{abstract}

\newcommand{\su}{{{ hu(1,1|sp(2)[M,2]) }}}
\newcommand{\hc}{{{ hc(1|2\!\!:\!\![M,2]) }}}
\newcommand{\hco}{{{ hc(1|(1,2)\!\!:\!\![M,2]) }}}
\newcommand{\hu}{{{ hu(1|2\!\!:\!\![M,2]) }}}
\newcommand{\hupm}{{{ hu^E_\pm(1|(1,2)\!\!:\!\![M,2]) }}}
\newcommand{\huo}{{{ hu(1|(1,2)\!\!:\!\![M,2]) }}}
\newcommand{\hue}{{{ hu^E(1|(1,2)\!\!:\!\![M,2]) }}}
\newcommand{\hunmpq}{{{ hu(n,m|(u,v,2)\!\!:\!\![M,2]) }}}
\newcommand{\hunm}{{{ hu(n,m|(0,1,2)\!\!:\!\![M,2]) }}}
\newcommand{\honm}{{{ ho(n,m|(0,1,2)\!\!:\!\![M,2]) }}}
\newcommand{\huspnm}{{{ husp(n,m|(0,1,2)\!\!:\!\![M,2]) }}}
\newcommand{\honmpm}{{{ ho_\pm(n,m|sp(2)|[M,2]) }}}
\newcommand{\huspnmpm}{{{ husp_\pm (n,m,|sp(2)|[M,2]) }}}
\newcommand{\hunmpm}{{{ hu_\pm(n,m|(0,1,2)\!\!:\!\![M,2]) }}}
\newcommand{\supm}{{{ hu_\pm(1,1|sp(2)[M,2]) }}}
\newcommand{\ty}{\hat{y}}
\newcommand{\bee}{\begin{eqnarray}}
\newcommand{\eee}{\end{eqnarray}}
\newcommand{\nn}{\nonumber}
\newcommand{\lis}{Fort1,FV1,LV}
\newcommand{\hy}{\hat{y}}
\newcommand{\by}{\bar{y}}
\newcommand{\bz}{\bar{z}}
\newcommand{\go}{\omega}
\newcommand{\e}{\epsilon}
\newcommand{\f}{\frac}
\newcommand{\p}{\partial}
\newcommand{\half}{\frac{1}{2}}
\newcommand{\ga}{\alpha}
\newcommand{\gal}{\alpha}
\newcommand{\U}{\Upsilon}
\newcommand{\C}{{\bf C}}
\newcommand{\Y}{{\mathcal Y}}
\newcommand{\ups}{\upsilon}
\newcommand{\bu}{\bar{\upsilon}}
\newcommand{\dga}{{\dot{\alpha}}}
\newcommand{\dgb}{{\dot{\beta}}}
\newcommand{\gb}{\beta}
\newcommand{\gga}{\gamma}
\newcommand{\gd}{\delta}
\newcommand{\gl}{\lambda}
\newcommand{\gk}{\kappa}
\newcommand{\gep}{\epsilon}
\newcommand{\gvep}{\varepsilon}
\newcommand{\gs}{\sigma}
\newcommand{\V}{|0\rangle}
\newcommand{\ws}{\wedge\star\,}
\newcommand{\gee}{\epsilon}
\newcommand\ul{\underline}
\newcommand\un{{\underline{n}}}
\newcommand\ull{{\underline{l}}}
\newcommand\um{{\underline{m}}}
\newcommand\ur{{\underline{r}}}
\newcommand\us{{\underline{s}}}
\newcommand\up{{\underline{p}}}
\newcommand\A{{\mathcal A}}
\newcommand\B{{\mathcal B}}
\newcommand\PP{{\mathcal P}}
\newcommand\M{{\mathcal M}}

\newcommand\uq{{\underline{q}}}
\newcommand\ri{{\cal R}}
\newcommand\punc{\multiput(134,25)(15,0){5}{\line(1,0){3}}}
\newcommand\runc{\multiput(149,40)(15,0){4}{\line(1,0){3}}}
\newcommand\tunc{\multiput(164,55)(15,0){3}{\line(1,0){3}}}
\newcommand\yunc{\multiput(179,70)(15,0){2}{\line(1,0){3}}}
\newcommand\uunc{\multiput(194,85)(15,0){1}{\line(1,0){3}}}
\newcommand\aunc{\multiput(-75,15)(0,15){1}{\line(0,1){3}}}
\newcommand\sunc{\multiput(-60,15)(0,15){2}{\line(0,1){3}}}
\newcommand\dunc{\multiput(-45,15)(0,15){3}{\line(0,1){3}}}
\newcommand\func{\multiput(-30,15)(0,15){4}{\line(0,1){3}}}
\newcommand\gunc{\multiput(-15,15)(0,15){5}{\line(0,1){3}}}
\newcommand\hunc{\multiput(0,15)(0,15){6}{\line(0,1){3}}}
\newcommand\ls{\!\!\!\!\!\!\!}

\newpage
\tableofcontents
\newpage

\section{Introduction}

\newcounter{lec}
\setcounter{lec}{1}

Aleksei Starobinsky was one of the world leaders in the field of quantum
gravity and cosmology (see, {\it e.g.,}  \cite{Sahni:1999gb}). He passed away tragically in a few days because
of the Covid 19 decease. I had a privilege to know Aleksei personally though
our scientific interests were different enough. Alexei was a deep knowledgable
person whose interests spread far beyond the scientific research. In this paper, dedicated to the memory of Aleksei, I wish to make a comment that may indicate
some convergence between our seemingly different fields.

The  aim of this letter is to point out that there exists an object (field) very different from the  fields usually used in  the Standard Model (SM), that may, in principle, affect general picture of the origin of some of phenomena in
high-energy physics, cosmology and astrophysics.    The main actor of this letter is the so-called singleton discovered by Dirac in 1963 in \cite{Dirac:1963ta}\footnote{The name was given even earlier in the math literature \cite{Ehrman}
to emphasize its simplicity from the representation theory perspective.}
as a specific branch of the solutions of a certain wave equation, that survives at infinity of $AdS_4$.
There are two singleton fields of boson and fermion types often called
$Rac$ and $Di$, respectively \cite{Flato:1978qz} (see \cite{Starinets:1998dt,Flato:1999yp} for reviews). Later on it was realized that $Rac$ and $Di$ and their supermultiplets considered in the context of supergravity $S^7$ compactifications, supermembrane and higher-spin  theory, for instance,
 in \cite{Sezgin:1983ik}-\cite{Duff:2025tot}
are nothing else as, respectively, free massless scalar and spinor fields in the 2+1 dimensional space presently identified with the boundary of $AdS_4$ within the paradigm of holographic correspondence \cite{Maldacena:1997re}-\cite{Witten:1998qj}, which interpretation is particularly relevant in the context of higher-spin holography \cite{Sezgin:2002rt,KP}.

In this letter we focus on the two issues.

 Firstly, we interpret singletons as  relativistic fields in $(A)dS_{d+1}$ providing
 for them Lorentz covariant field equations and actions which,
 to the best of our knowledge,  were not available before. This is achieved
 within the unfolded dynamics approach (see \cite{Vasiliev:2014vwa} for an elementary review and more references).
 The proposed formulation differs from the Flato-Fronsdal
 dipole one  \cite{Flato:1986uh} given  in terms of a certain fourth-order field equation  in $AdS_{d+1}$. In particular,  it is free from the huge gauge symmetry of \cite{Flato:1986uh} eliminating local degrees of freedom in $AdS_{d+1}$. On the other hand, kinematically  our formulation has some similarities with the  interpretation of holography proposed in the framework of the nonlinear realisation approach in \cite{Bellucci:2002ji,Ivanov:2003dt} (though not at the Lagrangian level).

Secondly, we speculate on the possible implications of the presence of the singleton
field in the context of SM and Gravity. Let us stress that our proposal
differs from seemingly analogous ideas of Flato and Fronsdal  engineering
usual fields, starting from the electromagnetic one \cite{Flato:1986bf}-\cite{Flato:1998iy} (and references therein), as composites of singletons. Instead we treat singletons democratically with all other fields of SM and Gravity as independent fields.

In particular, we  speculate that  singleton, that exists in presence of dark
energy, can be relevant to the dark matter problem  as well as  some other
physical phenomena including the
baryon asymmetry. It is important that though the genuine singleton is defined in $AdS_4$ our construction is applicable to $dS_4$ as well. Since singleton makes sense in presence of a nonzero cosmological
constant (=dark energy), this idea somehow unifies dark matter with
dark energy.

To put it short, our goal is to draw attention to the fact that,
apart from usual local relativistic fields, there may exist exotic
relativistic objects that are not local in the usual sense. Their characteristic feature
is that they form an infinite-dimensional representation of the Lorentz
group. A related phenomenon that singleton   cannot be
localised at a point in the $3d$ space of $4d$ space-time is  known from
the very first singleton papers \cite{Flato:1986uh,Flato:1999yp}. As a result, from
the $4d$ space-time perspective  it is nowhere (= everywhere).
This means in particular that singleton does not affect the local scattering and radiation SM processes. On the other hand, that singletons form representations of the Lorentz algebra makes it possible to introduce their gravitational interaction within  Cartan formalism. As argued in the paper, singleton can contribute to collective physical phenomena beyond the standard collider physics.

The paper is organized as follows. In Section \ref{Unfolded Dynamics} we
sketch main ideas of the unfolded dynamics approach underlying our construction.
 The  mechanism allowing to formulate the same dynamics in spaces of different dimensions is recollected in Section \ref{meta}.
Unfolded formulation of the conformal scalar and spinor in $d$ dimensions is
presented in Sections \ref{csca} and \ref{cspi}, respectively, at the  both off-shell and on-shell levels, including conformal invariant Lagrangians. Extension of the singletons to $(A)dS_{d+1}$  is presented in Section \ref{ext}. Spinor formulation of the on-shell $3d$ singletons and their $4d$ extensions is presented
 in Section \ref{3d}. In Section \ref{BSM} we speculate on the potential role of  singletons for the resolution of some  problems in SM and related aspects of cosmology and astrophysics. Main results and further research directions are summarised in Section \ref{conc}.

\section{Unfolded dynamics}
\label{Unfolded Dynamics}
{The unfolded form of dynamical equations is a generalization
of the first-order form of ordinary  differential equations}
\be
\dot{q}^i(t) =\varphi^i (q(t))\,
\ee
resulting from the
 replacement of the time derivative by the de Rham derivation,
$$\f{\p}{\p t} \to \dr= \theta^n \p_n\q \theta^n \theta^m = -\theta^m \theta^n\q dx^n\equiv \theta^n\,,
$$
and the dynamical variables $q^i$ by a set of differential forms $W^\Omega (\theta,x)$
$$
 q^i(t)\rightarrow W^\Omega (\theta,x)=\sum_{p=0}\theta^{n_1}\ldots \theta^{n_p}
W^\Omega_{n_1 \ldots n_p} (x)\,, $$
that allows one
to reformulate a system of partial differential equations in the
first-order  form
\be
\label{unf}
\dr W^\Omega (\theta,x)=G^\Omega (W(\theta,x))\,.\qquad
\ee
Here
$G^\Omega (W)$ {are some functions of the ``supercoordinates"}
$W^\Omega$,
$$
G^\Omega (W) = \sum_n f^\Omega{}_{\Lambda_1\ldots
\Lambda_n}W^{\Lambda_1}  \ldots W^{\Lambda_n}\,.
$$
Since $\dr^2 =0$, at $d>1$ {the functions $G^\Lambda (W)$  have to obey
the  compatibility conditions}
\be
\label{cc}
 G^\Lambda (W)\f{\p
G^\Omega (W)} {\p W^\Lambda}  \equiv 0\,.
\ee
Let us stress that these are conditions on the functions $G^\Lambda (W)$ rather than on $W^\Omega$.

The idea of the unfolded formulation was put forward in
\cite{Vasiliev:1988xc} where it was realized that the full system of
nonlinear equations for massless higher-spin gauge fields can be searched in the form  (\ref{unf}) as a
deformation of the free unfolded equations.

As a consequence of the compatibility conditions (\ref{cc}) the system (\ref{unf}) is
invariant under the gauge transformation
\be
\label{delw} \delta W^\Omega = \dr \varepsilon^\Omega +\varepsilon^\Lambda
\frac{\p G^\Omega (W) }{\p W^\Lambda}\,,
\ee
 {where the gauge parameter} $\varepsilon^\Omega (x) $ {is a}
$(p_\Omega -1)$-form if $W^\Omega$ was a $p_\Omega$-{form}. {Strictly
speaking, this is true for the class of {\it universal} unfolded systems in which
the compatibility conditions (\ref{cc}) hold independently of the dimension  $d$ of
space-time, \ie (\ref{cc}) should be
true disregarding the fact that any $(d+1)$-form is zero. This is the case in
all unfolded systems considered in the higher-spin literature including this paper.}

As shown in \cite{Vasiliev:2005zu}, the variety of invariant functionals
associated with the unfolded equations (\ref{unf}) is described by the cohomology
of the operator
\be
\label{Q}
Q= G^\Omega \f{\p}{\p W^\Omega}\,,
\ee
that obeys
\be
Q^2=0
\ee
as a consequence of (\ref{cc}). As a result, the unfolded equations can be written
in the Hamiltonian-like form
\be
\label{unf1}
\dr F(W) = Q(F(W))\q \forall F(W)\,.
\ee

By virtue of (\ref{unf1}), $Q$-closed $p$-forms $L_p(W)$ are $\dr$-closed, giving rise to the gauge invariant functionals
\be
S=\int_{\Sigma^p} L_p\,.
\ee
(See \cite{Vasiliev:2005zu} for more detail and examples.)

A particular example of an unfolded system is provided by the zero-curvature (Maurer-Cartan)
equations. Namely, let $h$ be a Lie algebra with the basis
$\{T_\a\}$. Let $\o=\o^\a T_\a$ be a one-form
valued in $h$. For $G (\,\o)=-\o\o:= -\frac{1}{2}
\o^\a \o^\b [T_\a , T_\b]$ equation (\ref{unf}) with
$W=\o$ reads as\footnote{The exterior product wedge symbol is omitted in this paper since all products are
automatically exterior due to the presence of anticommuting $\theta$.}
\be
\label{0cur} \dr\o+\o
 \o=0\,.
\ee
 Relations (\ref{cc}) and (\ref{delw}) amount, respectively,
to the usual Jacobi identity for $h$ and gauge
transformation for  $\o$.

If the set $W^\a$  contains some $p$-forms $\cc^i$
({\it e.g.} zero-forms) and  the functions $G^i$ are linear in $\o$
and $\cc$,
\be
\label{lin}
 G^i =- \o^\a(T_\a)^i {}_j  \cc^j\,,
\ee
 then
(\ref{cc}) implies that the matrices $(T_\a)^i {}_j$ form some
representation $T$ of $h$, acting in a space $V$ where $\cc^i$
is valued. The associated equation (\ref{unf}) is a
covariant constancy condition \be \label{covc} D_\o \cc=0 \ee
 with $D_\o\equiv \dr+\o$ being a  covariant derivative in the
$h$-module $V$.

The zero-curvature equations (\ref{0cur}) usually describe
background geometry in a coordinate independent fashion.
For instance, let $h$ be the Poincar\'{e} algebra with the gauge fields
\be
\o (x)=e^n (x) P_n
+\half \o^{nm} (x) M_{nm}\,,
\ee
where $P_n$ and $L_{nm}$ are generators of
translations and Lorentz transformations with $e^n (x)$ and
$\o^{nm}(x)$  identified with the frame one-form
and Lorentz connection, respectively
(fiber Lorentz vector indices $m,n\ldots$ run from 0 to $d-1$
and are raised and lowered by the flat Minkowski metric).
It is well known that the zero-curvature condition (\ref{0cur}) for the
 Poincar\'{e} algebra describes Minkowski geometry in a coordinate-independent
way.

By choosing a different Lie algebra $h$ one can describe
a different background like, {\it e.g.}, anti-de Sitter for $h=o(d-1,2)$
or conformally flat for $h=o(d,2)$. The covariant constancy
equation (\ref{covc}) then describes $h$-invariant linear equations in a chosen
background.

The unfolded formulation  has a number of
remarkable properties starting from its general applicability:
 every system of partial differential equations can be
reformulated in the unfolded form.
{Due to  the exterior algebra formalism, the system
is  coordinate independent.}

Local degrees of freedom are represented by the subset of zero-forms $C^I(x_0)\in \{W^\Omega(x_0)\}$
{at any} {$x=x_0$}. This is analogous to the fact that $q^i(t_0)$
describe degrees of freedom in the first-order form of ordinary differential
equations. In field-theoretic models,
the zero-forms $C^I(x_0)$ realize  {an infinite-dimensional module dual to the space
of single-}{particle states of the system in question} \cite{Vasiliev:2001zy}. This space is
analogous  to the phase space  in the Hamiltonian dynamics.

\section{Space-time metamorphoses and holography}
\label{meta}
{Unfolded dynamics exhibits independence of the ``world-volume" space-time with
coordinates $x$.}
{Instead, geometry is encoded by} the functions $G^\Omega (W)$ in the ``target space" with fields $W^\Omega$ as local coordinates.
Since the universal unfolded equations make sense in any space-time independently of
a particular realization of the de Rham derivation $\dr$, one is free to
extend space-time by  additional coordinates $z^u$,
\be
 \dr W^\Omega (x)=G^\Omega (W(x))\,,\quad x\rightarrow X=(x,z)\,,\quad
\dr_x\rightarrow \dr_X = \dr_x +\dr_z\,,\quad \dr_z =
dz^u\f{\p}{\p z^u}\,.
\ee
Locally,  unfolded equations reconstruct the
$X${-dependence in terms of values of the fields} $W^\Omega(X_0)=W^\Omega(x_0,z_0)$
{at any} $X_0$. {Clearly, to take} $W^\Omega(x_0,z_0)$ in space $M_X$
{with coordinates} $X_0$ {is the same as to take} $W^\Omega(x_0)$ {in the space}
$M_x\subset M_X$  with coordinates  $x$.

{Generally,  unfolding can be interpreted  as some sort of  covariant twistor
transform} \cite{Vasiliev:2014vwa,Misuna:2024dlx}
\bee\label{diapen}\nn
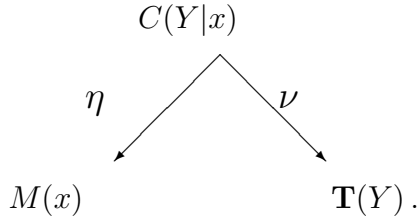
 \begin{picture}(200,80)( 0,26)
{\linethickness{.25mm}
\put(90,90){\vector( 1,-1){40}}%
\put(90,90){\vector( -1,-1){40}}%
\put(55,100)  {{    $C(Y|x)$}}
\put(6,32)  {{   $M(x)$}}
\put(128,32)  {{   $\mathbf{T}(Y)\,.$}}
\put(35,70)  {{  \large $\eta$}}
\put(108,70)  {{  \large $\nu$}}
 }
\end{picture}
\eee
Here
 $W(Y|x)$  {are functions on the
``correspondence space"} $C$ with  local coordinates $Y,x$.
{The space-time} $M$ {has local coordinates} $x$.
 {The twistor space}
 $\mathbf{T}$ {has local coordinates} $Y$ the expansion over which generates the components $W^\Omega(x)$ with various $\Omega$  (for examples see Sections \ref{csca}, \ref{cspi}, \ref{3d}).

 {Unfolded equations reconstruct the dependence of $W(Y|x)$ on $x$
 in terms of the function $W(Y|x_0)$ on $\mathbf{T}$  at some  $x_0$.
 The restriction of $W(Y|x)$ or some its $Y$-derivatives to $Y=0$
 gives dynamical fields $\go(x)$ in $M$ which, in the on-shell case,
  solve their dynamical field equations.
 Hence, similarly to the Penrose transform (see \cite{Vasiliev:2014vwa}
 and references therein), unfolded equations  map functions
 on} $\mathbf{T}$ {to solutions of the dynamical field equations in} $M$.

{In these terms, the holographic duality can be interpreted as the duality between
different space-times $M$ that can be associated with the same twistor space.}
The problem becomes most interesting
provided that there is a nontrivial vacuum connection along the additional
coordinates $z$. This is in particular the case of $AdS/CFT$ correspondence where
the conformal flat connection at the boundary is extended to the flat $AdS$
connection in the bulk with $z$ being a Poincar\'e  coordinate.
This mechanism has a number of interesting applications. In particular, in
\cite{Vasiliev:2001zy}
it was applied to the description of all $4d$  massless fields in terms of a single
scalar in the ten-dimensional space identified by Fronsdal with Lagrangian Grassmannian in \cite{lag} (see also \cite{Bandos:1999qf,Bandos:2005mb}).

{Conventional holography \cite{Maldacena:1997re}-\cite{Witten:1998qj} is based on the isomorphism of the boundary conformal group} $O(d,2)$ {with  the symmetry of} $AdS_{d+1}$.
Generators of the conformal algebra obey the relations
\be
\label{dp}
[D\,, P_a] = -P_a\q [D\,, K^b] =  K^b\q [D\,,L_{ab}]=0\,,
\ee
\be
[P_a\,, K_b] = 2L_{ab} -2 \eta_{ab} D\,,
\ee
\be
[L_{ab} \,,P_c ] = \eta_{bc} P_a - \eta_{ac} P_b\q
[L_{ab} \,,K_c ] = \eta_{bc} K_a - \eta_{ac} K_b\,,
\ee
\be
\label{LL}
[L_{ab} \,,L_{cd} ] = \eta_{bc} L_{ad} - \eta_{ac} L_{bd}-\eta_{bd} L_{ac} + \eta_{ad} L_{bc}\,.
\ee

Let $M^d$ be a $d$--dimensional conformally flat space-time
with local coordinates $\xx$ and  some $o(d,2)$
connection  $w_{\xx}(\xx)=
w^A_{\xx} T_A$ obeying the flatness conditions
\be
\label{drxgo}
\dr_\xx w_{\xx}(\xx) + w_{\xx}(\xx)w_{\xx}(\xx) =0 \q \dr_\xx := d\xx^n \f{\p}{\p \xx^n}\,.
\ee
A particular flat connection,
that corresponds to Cartesian coordinates in $M^d$, is
\be
\label{Cart}
w_{\xx}(\xx)= d\xx^a P_a\,.
\ee

The dilatation generator $D$ induces standard $\mathbb{Z}$ grading
on $o(d,2)$,
\be
[D\,,T_A ]= \Delta (T_A) T_A\,
\ee
with $\Delta(T_A)$ being the conformal dimension of $T_A$,
\be
\label{D}
\Delta(L)=0\q\Delta(D)=0\q \Delta(K) = 1\q \Delta (P) = -1\,.
\ee

Let us now introduce an additional coordinate $z$ and differential $dz$
so that $x=(\xx,z)$ be local coordinates of $AdS_{d+1}$. A
conformally foliated connection $W(x)$ of $AdS_{d+1}$ can be
introduced as follows. The components of the connection
with differentials $d\xx$ are
\be
\label{fol}
W_{\xx}^A(x) T_A = z^{\Delta(T_A)} w^A_{\xx}(\xx) T_A\,,
\ee
while the only nonzero $dz$ component of the connection
is associated with the dilatation generator $D$, having the form
\be
W_{z}(x) D =  - z^{-1} dz D \,.
\ee
Clearly, so defined connection $W(x)$ is flat in
(a local chart of) $AdS_{d+1}$.
Poincar{\'e}
coordinates result from this construction applied to the connection (\ref{Cart}).

Analogously, unfolded equations
\be
\D_\xx C_i(\xx)=0\q \D_\xx := \dr_\xx + W^A_\xx  T_A\
\ee
in $M^d$ for a set of fields $C_i(\xx)$ carrying
conformal weights $\Delta_i$ extend to the fields
\be
\label{zexx}
C_i(x) = z^{\Delta_i} C_i (\xx)\,
\ee
and equations
\be
\label{eqx}
\D_x  C_i(x)=0\q \D_x := \dr_x + W^A_x  T_A\,.
\ee

It is important to note that if a system was off-shell in $M^d$ this is not
so in the extended $d+1$-dimensional space. Indeed, the dependence on the
additional coordinate $z$ is determined by (\ref{eqx}) in terms of that on
$\xx$ as is most obvious from (\ref{zexx}). This means that the field in $AdS_{d+1}$
obeys some differential equation, that determines its $z$-dependence.

To identify the $d+1$-dimensional space with (a local chart of) $AdS_{d+1}$ it suffices to redefine $o(d,2)$ generators as
\be
\label{AdSgen}
 \PP_\nu = \big ( ( P_a+\lambda^2 K_a) \,, 2\lambda D\big )\q \M_{\nu\mu} =\big (L_{ab}, \f{1}{2\lambda} (P_a-\lambda^2 K_a)\delta_\nu^d, -
\f{1}{2\lambda} (P_b-\lambda^2 K_b)\delta_\mu^d \big )
\ee
with $a,b = (0\,,\ldots ,d-1 )$ $\nu,\mu = (0\,,\ldots , d )$, interpreting $\PP_\nu$
and $\M_{\nu\mu}$ as $AdS_{d+1}$ translation (transvection) and Lorentz generators,
respectively. The respective connections are
\be
\label{WADS}
W = h^\nu \PP_\nu +\half \go^{\nu\mu} \M_{\nu\mu}\,.
\ee
 In particular, this means that
\be
\label{ehM}
E^a = \half(h^a + 2 \lambda \go^{ad})\,,
\ee
where $E^a$ is the $d$-dimensional vielbein rescaled in accordance with (\ref{fol}) (the index $d$ in (\ref{ehM}) indicates the direction along  $z$).
The dimensionful parameter $\lambda$ is related to the cosmological constant
$\Lambda$,
\be
\Lambda =-\#\lambda^2\,
\ee
with   some positive number $\#$.
This implies that $\lambda$ is real and pure imaginary  in the $AdS_{d+1}$ and $dS_{d+1}$ cases, respectively.
Naively, this suggests that the construction is not working
in the de Sitter space. In fact, this is not necessarily true and, moreover, as discussed below, specificities of the $dS$ case
 may be of relevance to the baryon asymmetry problem.

According to the Flato-Fronsdal theorem \cite{Flato:1978qz},
{boundary conformal currents are dual to fields in the bulk} $AdS_{d+1}$. In other words, free relativistic fields in the bulk are associated with the bilinear
currents on the boundary, the fact underlying the
Klebanov-Polyakov HS holographic conjecture \cite{KP}.
{Here we consider an opposite situation: starting
 from the boundary conformal field we will
 see what is its bulk dual. Our construction  differs from the
 other holographic treatments of singletons (see e.g.
 \cite{Starinets:1998dt,Kogan:1999bn,Ohl:2012bk})
 based on the  dipole
 singleton description of \cite{Flato:1986uh}.
 The output is  interesting both formally
 and, hopefully,
 physicswise shedding  more light on
  what are bulk duals of the free
 conformal boundary fields, \ie singletons.

\section{Conformal scalar in any $d$ within unfolded formalism}
\label{csca}

Singleton $Rac$ is a massless conformal scalar field in any dimension $d$. In the unfolded
dynamics approach it is described as follows \cite{Shaynkman:2000ts}. Let $C(y|\xx)$
be a zero-form, that depends on the space-time coordinates $\xx^n$ and auxiliary
variables $y^n$ ($n=0,\ldots d-1$). Consider unfolded equations of the form
\be
\label{off}
\dr_\xx C (y|\xx) + d\xx^n \f{\p}{\p y^n} C (y|\xx) =0\,.
\ee
Clearly, this equation relates the coefficients $C_{a_1\ldots a_n}(\xx) $
 of the expansion
\be
\label{C}
C (y|\xx) = \sum_{n=0}^\infty \f{1}{n!} C_{a_1\ldots a_n}(\xx) y^{a_1}\ldots y^{a_n}
\ee
to higher $\xx^a$ derivatives of the ground component  $C(\xx)$,
\be
\label{der}
C_{a_1\ldots a_n}(\xx) = (-1)^n \p_{a_1}\ldots \p_{a_n} C(\xx) \q \p_a:=\f{\p}{\p \xx^a}\,,
\ee
\be
C(\xx):= C(0|\xx)\,
\ee
identified with the scalar field. The system (\ref{off}) is off-shell,
imposing no
differential conditions on $C(\xx)$. To put the system on shell of a massless field
it suffices to   constrain $C (y|\xx)$ by the condition
\be
\label{boxy}
\Box_y C(y|\xx) =0\q \Box_y := \eta^{ab} \f{\p^2}{\p y^a \p y^b}\,.
\ee

The system (\ref{off}) is equivalent to
\be
\label{dc}
\D_\xx C(y|\xx) =0\,,
\ee
where
\be
\label{D}
\D_\xx:= \dr_\xx + e^a P_a  + f_a K^a + \half \go^{ab} L_{ab} + b D
\ee
is the covariant derivative of the conformal  algebra $o(d,2)$ with the
generators $P_a$ for translations, $K^a$ for special conformal transformations, $L_{ab}$ for Lorentz transformations and $D$ for dilatations. The particular flat connection used in (\ref{off}) is
\be
\label{Cart1}
e^a = d \xx^a\q \go^{ab} =0\q b =0\q f_a=0\,.
\ee

In terms of $y^a$,
the conformal generators, that obey
(\ref{dp})-(\ref{LL}), are realized as
\be
\label{para}
P_a =\f{\p}{\p y^a}\q L_{ab} = y_a \f{\p}{\p y^b} - y_b \f{\p}{\p y^a}\q
D= y^a\f{\p}{\p y^a} +\Delta \,,
\ee
\be
\label{Ky}
K_a = y^2 \f{\p}{\p y^a} - 2y_a y^b \f{\p}{\p y^b}  - 2 \Delta y_a\,,
\ee
where  $\Delta$ is a number (conformal weight). The system (\ref{dc}) along with
the consistency condition
\be
\D_\xx^2=0
\ee
equivalent to the flatness condition (\ref{drxgo}) forms an unfolded
system invariant under the gauge conformal transformations (\ref{delw}). Any choice of a particular flat connection $e^a$, $f_a$, $\omega_{ab}$, $b$ in (\ref{D}) restricts the local conformal transformations (\ref{delw}) to the global ones. (Analogously the choice of Minkowski metric restricts diffeomorphisms to global Poincar\'{e} transformations. For more detail see
\cite{Vasiliev:2014vwa}).
This proves global conformal invariance of the system (\ref{dc}).

Note that the construction of this section
is extendable to $(A)dS_d$ as well as to any other conformally flat background
by the appropriate choice of the flat connection of the conformal group.
In particular, conformal field theories in $AdS_d$ and their holographic aspects  were discussed in
\cite{Vasiliev:2001zy,Aharony:2010ay,Nilsson:2012ky,Ohl:2012bk}.

Next one observes that if $C(Y|\xx)$ obeys the constraint (\ref{boxy}), it is obeyed by
$T_A C(Y|\xx)$ for all conformal algebra generators $T_A$  provided that
\be
\label{deltacan}
\Delta = \frac{d}{2} -1\,,
\ee
which is the canonical dimension of a massless scalar in $d$ dimensions.
 This implies that the constraint (\ref{boxy}) is preserved by the equation
(\ref{dc}). Since the system (\ref{boxy}), (\ref{dc}) is equivalent to the massless
Klein-Gordon equation, this in turn proves conformal invariance of the latter.

From the representation theory perspective this implies that the module $V^{tr}$ generated by the conformal generators from a constant in $y^a$ only contains $y$--traceless polynomials  $f(y)$ obeying $\Box_y f(y)=0\,.$ To see this it is instructive to check that $K^a K_a$ applied to a constant yields zero for $K^a$ (\ref{Ky}) and
$\Delta$ (\ref{deltacan}). Hence, $V^{tr}$ is a submodule of the module
$V$ of all polynomials,  $V^{tr}\subset V$, while traceful polynomials in $y^a$
are in the factor module $V/V^{tr}$.

Now we observe that the coefficients $C_{a_1\ldots a_n}(\xx) $ in the expansion (\ref{C}) are in the dual
module $V^*$ with respect to the action of the covariant derivative $\D$  (\ref{dc}), (\ref{D}).
This means that $ (V/V^{tr})^*$ is a submodule of $V^*$ while  $(V^{tr})^*$ is a factor-module.
As a result, there are two components of $C(y)$ the covariant derivative of which does not contain the gauge field $f^a$ of special conformal transformations.
One is the vacuum (lowest weight) component $C(\xx)$, while  another is the
singular vector (synonymous to be annihilated by $K^a$) associated with the trace component
\be\label{sing}
C'(\xx):=C^a{}_a(\xx)\,.
\ee
This implies that to prove conformal invariance of any functional built from $C(\xx)$ and $C'(\xx)$ it suffices to check its
invariance under the action of the parabolic subalgebra generated by $P_a$, $L_{ab}$ and $D$. (For more detail on the derivation of general conformal invariant equations
from the representation theory of the conformal group see \cite{Shaynkman:2004vu} and references therein.)

The conformal invariant Lagrangian for a scalar field is a $d$-form
\be
\label{L}
L^{Rac} = \half \epsilon_{a_1\ldots a_d} e^{a_1}(\xx) \ldots  e^{a_d}(\xx) C(\xx) C'(\xx)\,,
\ee
where $e^a$ is a vielbein one-form in $d$ dimensions.
It is easy to see that $L^{Rac}$ is $Q$-closed with respect to $Q$ (\ref{Q}). Indeed, the special conformal
gauge field does not appear in $QL$ since it is absent in $de^a$, $\D (C)$ and $\D(C')$. Lorentz connection $\omega^{ab}$
cancels because $L$ is Lorentz invariant. Analogously, $b$ cancels because the Lagrangian has proper scaling dimension due to (\ref{deltacan}).
Finally, the contribution of $e^b$ cancels because of antisymmetrization over $d+1$ indices $a=0,1,\ldots d-1$ due to the exterior product (\ie $\theta$--dependence) of the one-forms $e^a$.

The fields $C(y|\xx)$ still obey unfolded equations (\ref{dc}), that are off-shell   just expressing higher components $C_{a_1\ldots a_n}(\xx)$ via derivatives of $C(\xx)$ according to (\ref{der}), imposing no
differential conditions on the latter. In particular, they imply that $C'(\xx)=\Box_\xx C(\xx)$. The Lagrangian field equations then imply the massless equation $\Box_\xx C(x)=0$.

\section{Conformal spinor in any $d$ within unfolded formalism}
\label{cspi}

Conformal spinor $Di$ is described analogously to  scalar $Rac$.
Consider unfolded equations of the form
\be
\label{offs}
\dr_\xx C_\ga  (y|\xx) + d\xx^a \f{\p}{\p y^a} C_\ga (y|\xx) =0\,
\ee
with a spinor index $\ga$.
This equation relates the coefficients $C_{\ga, a_1\ldots a_n}(\xx) $
of the expansion
\be
\label{Csp}
C_\ga (y|\xx) = \sum_{n=0}^\infty \f{1}{n!} C_{\ga,a_1\ldots a_n}(\xx) y^{a_1}\ldots y^{a_n}
\ee
to higher derivatives in $\xx^a$,
\be
\label{derf}
C_{\ga, a_1\ldots a_n}(\xx) = (-1)^n \p_{a_1}\ldots \p_{a_n} C_\ga (\xx) \,,
\ee
where $C_\ga (\xx)$ is the ground component of $C_\ga (y|\xx)$,
\be
C_\ga (\xx):= C_\ga (0|\xx)\,
\ee
identified with the genuine spinor field. As such, the system is off-shell, imposing no differential conditions on $C_\ga (\xx)$.

To put the system on shell of a massless field it suffices to impose the  constraint on $C (y|\xx)$
\be
\label{ggay}
\gga^a{}_\ga{}^\gb \f{\p}{\p y^a} C_\gb (y|\xx) =0\,,
\ee
where $\gga^a$ are gamma-matrices,
\be
[\gga^a \,,\gga^b ]=2\eta^{ab} Id\,.
\ee
The system (\ref{offs}) is the particular case of
\be
\label{dcs}
\D_\xx C_\ga (y|\xx) =0\,
\ee
for $\D_\xx$ of the form (\ref{D}), (\ref{Cart1})
with the conformal generators
\be
\label{paras}
P_a =\f{\p}{\p y^a}\q
L_{ab} = y_a \f{\p}{\p y^b} - y_b \f{\p}{\p y^a}+\f{1}{4} [\gga_a\,,\gga_b] \q
D= y^a\f{\p}{\p y^a} +\Delta \,,
\ee
\be
\label{Kys}
K_a = y^2 \f{\p}{\p y^a} - 2y_a y^b \f{\p}{\p y^b}  - 2 \Delta y_a +
\half y^b [\gga_b\,,\gga_a] \,.
\ee

It is not hard to see that, for
\be
\label{canspin}
\Delta = \f{d-1}{2}\,,
\ee
which is a canonical dimension of the massless spinor, the action of $K^a \gga_a$
on a $y^a$-independent element yields zero.
This implies that $\gga^a$--transversal polynomials form a submodule
$V^{\gga tr}$ of $V^{sp}$ of all polynomials (\ref{Csp}).
Analogously to the scalar case, this has a consequence that the space
of spinor fields
(\ref{Csp}) possesses a lowest weight vector $C_\ga (\xx)$
identified with the dynamical spinor field and a singular vector $C'_\ga (\xx)$ associated with the Dirac operator,
\be
C_\ga (\xx):=C_{\ga}(0|\xx) \q C'_\ga (\xx) := \gga^a{}_{\ga}{}^\gb C_{\gb a}(0|\xx)\,.
\ee
The conformally invariant Lagrangian has the form analogous to (\ref{L}),
\be
\label{Lsp}
L^{Di} = \half \epsilon_{a_1\ldots a_d} e^{a_1}(\xx) \ldots e^{a_d}(\xx) \bar C^\ga (\xx) C'_\ga (\xx)\,,
\ee
where $\bar C^\ga (\xx)$ is the Dirac conjugated spinor. Clearly, in the Cartesian coordinate system (\ref{Cart1}), Eq.~(\ref{Lsp}) yields  usual
Dirac Lagrangian for the spinor field $C_\ga(\xx)$ in $d$-dimensions.

\section{Extension to $(A)dS_{d+1}$}
\label{ext}
Now we are in a position to extend the $d$-dimensional  singleton systems to   $AdS_{d+1}$. To this
end we replace equations (\ref{dc}), (\ref{dcs}) by analogous equations
\be
\label{unfads}
D_x C(x) =0\q D_x C_{\ga} (x)=0\q D_x := \dr_x + W
\ee
 with $W$ (\ref{WADS}). The conformal generators
have the form (\ref{para}), (\ref{Ky}) and (\ref{paras}), (\ref{Kys}) in the scalar and spinor cases, respectively.

The $AdS_{d+1}$ invariant Lagrangians still have the form (\ref{L}) for scalar
and (\ref{Lsp}) for spinor but now
being  $d$-forms in $(A)dS_{d+1}$ with the fields  $C$ and $C'$ rescaled by (\ref{zexx}) and $E^a$ (\ref{ehM}),
\be
\label{Lx}
L^{Rac} = \half \epsilon_{a_1\ldots a_d} E^{a_1}(x) \ldots  E^{a_d}(x)
 C(x) C'(x)\,,
\ee
\be
\label{Dix}
L^{Di} = \half \epsilon_{a_1\ldots a_d} E^{a_1}(x) \ldots  E^{a_d}(x)
 \bar C^\ga (x) C'_\ga(x)\,.
\ee

These Lagrangians are closed and, as a consequence of the general properties of the unfolded equations,  invariant up to exact forms (\ie total derivatives) under the symmetries
(\ref{delw}) that leave invariant the background connections, that is global $(A)dS$ symmetries.
However, now one has to take into account that equations (\ref{unfads}) are no longer off-shell reconstructing the dependence of one of the coordinates, namely $z$, in terms of the others. This is the reason why the seemingly non-invariant form
of (\ref{Lx}), (\ref{Dix}) in view of (\ref{ehM}) still respects Lorentz covariance in $AdS_{d+1}$. In other words, the $d+1$-dimensional Lorentz transformations act on the singleton
nonlocally relating fields $\phi(\xx,z)$ at different $z$.
As a result, being a local field in $d$ dimensions, from the $d+1$ perspective it is nowhere (equivalently, everywhere).

As mentioned in Section \ref{meta}, the parameter $\lambda$, that enters the $(A)dS$ connection by virtue of (\ref{AdSgen}), (\ref{WADS}), is real in the $AdS$ case
but pure imaginary in  $dS$. Naively, this implies that the Lagrangians
(\ref{Lx}) and (\ref{Dix}) are not Hermitian in the $dS$ case. However, this problem
can be resolved by introducing doublets of mutually conjugated  fields $C^\pm$ and/or $C^\pm_\ga$, associated with $\lambda =\pm i \lambda'$ with real $\lambda'$. This allows one to consider singletons as fields in the de Sitter space
tantamount to dark energy \cite{Sahni:1999gb}. The modes associated with the evolution along $z$ are
either increasing or decreasing that is not too surprising  in the expansion regime.

\section{Spinor formulation for $3d$ singletons}
\label{3d}
In 2+1 dimensions it is convenient to describe massless fields in terms of
spinor indices  $\ga,\gb =1,2$. In these terms, local coordinates are
$\xx^{\ga\gb}=\xx^{\gb\ga}$ and massless field equations are
\be
\f{\p}{\p \xx^{\ga \gb}} \f{\p}{\p \xx_{\ga\gb}} \phi(\xx)=0
\ee
 and
\be
\f{\p}{\p \xx^{\ga \gb}} \phi^\gb (\xx)=0
\ee
for scalar $\phi(\xx)$  and spinor $\phi^\gb (\xx)$, respectively. The unfolded massless field equations are formulated  \cite{Shaynkman:2001ip} both for $Rac$ and for $Di$
with the aid of auxiliary spinor variables $y^+_\ga$, $y^-_\gb$ obeying
$$
[y^-_\ga \,, y^+_\gb ] = \gvep_{\ga\gb}
$$
in terms of the Fock module
\be
\label{Fock}
|\phi (y^+ |\xx)\rangle =\phi(y^+ |\xx) |0\rangle\q y^{-}_\ga  |0\rangle =0\q
\phi(y^+ |x) = \sum_{n=0}^\infty \f{1}{n!} \phi_{\ga_1\ldots \ga_n}(x)
y^{+\ga_1}\ldots y^{+\ga_n}\,
\ee
in the form
\be
D_\xx |\phi (y^+ |\xx)\rangle =0
\ee
with the $o(3,2)\sim sp(4|\mathbb{R})$ covariant derivative
\be
D_\xx:= \dr_\xx + e^{\ga\gb}(\xx) y^{-\ga} y^{-\gb} + \go^{\ga\gb}(\xx) y^+_\ga y^-_{\gb} +
b(\xx)\{ y^{+\ga}\,, y^-_{\ga}\} +  f(\xx)_{\ga\gb} y^{+\ga} y^{+\gb}\,
\ee
for some flat connection obeying $D_\xx^2=0$. Cartesian coordinate system results from the  flat  connection
\be
 e^{\ga\gb}(\xx) = d\xx^{\ga\gb}\q \go^{\ga\gb}=0\q b=0\q f_{\ga\gb}=0\,.
\ee

To describe singleton as a $4d$ field  it suffices to add an additional coordinate and extend the
flat $o(3,2)$ connection to the four-dimensional space.
Identifying the additional coordinate with the Poincar\'{e} coordinate $z$ we set
\be\nn
\label{xz}
\xx^{\ga\gb}\to
x^{\ga\dga }=(\xx^{\ga\dga},-\f{i}{2} \epsilon^{\ga\dga} z^{-1})\,.
\ee

The unfolded field equations in $AdS_4$ read as
\be
\label{dx}
D_x |\phi\rangle =0\,
\ee
with
$$
D_x = \dr_x +\f{i}{ z}d\xx^{\ga\gb} y^-_\ga y^-_\gb - \f{dz}{2z} y^-_\ga
y^{+\ga}\q \dr_x:= dx^{\ga\dgb} \f{\p}{\p x^{\ga\dgb}}
$$
describing a flat $AdS_4$ connection   in Poincar\'e coordinates with
\be\label{poframe}\nn
e^{\ga\dga} = \f{1}{2z} dx^{\ga\dga}\q \go^{\ga\gb}=-
\f{i}{4z} d\xx^{\ga\gb}\q \bar \go^{\dga\dgb} =
\f{i}{4z} d\xx^{\dga\dgb}\,.
\ee
(For more detail see \cite{Vasiliev:2012vf}.)

Though unfolded equations are formulated in the $4d$ space-time, the dynamical equations are still three-dimensional}
$$
\f{\p^2}{\p \xx^{\ga\gb}\p \xx_{\ga\gb}} \phi (\xx,z)=0
$$
at every $z$. The equations (\ref{dx}) then reconstruct $z$ dependence of
$\phi(\xx,z)$ in terms of $\phi(\xx,z_0)$ at any $z_0\neq 0$.
The system is relativistic in the $4d$ sense since the Lorentz algebra $o(3,1)$ is a subalgebra of $o(3,2)$ as well as of $o(4,1)$. However, the singleton field treated as a relativistic field in the $4d$
space-time is very different from usual local space-time fields since it belongs to
an infinite-dimensional $o(3,1)$ module, namely, the Fock module (\ref{Fock}).
This is because the $4d$ Lorentz generators $L^{\ga\gb}$ and $\bar L^{\dga\dgb}$ realized as
\be
L_{\ga\gb} = \half (y^+_\ga +i y^-_\ga)(y^+_\gb +i y^-_\gb)\q
\bar L_{\dga\dgb} = \half (y^-_\dga +i y^+_\dga)(y^-_\dgb +i y^+_\dgb)
\ee
(see \cite{Vasiliev:2012vf}) contain the creation $y^+ y^+$ parts.

To formulate an off-shell Lagrangian description of the $3d$ singleton
in terms of spinors one has to introduce along the lines of the $4d$
construction of \cite{Misuna:2020fck} an additional scalar variable $p$
parameterising the (gamma)traceful components of the off-shell $3d$ fields
$(C_\ga(\xx)) C(\xx)$. The off-shell extension will be elaborated elsewhere.

\section{Singleton as a beyond SM  actor}
\label{BSM}

That $4d$ Lorentz algebra acts on the singleton is crucially important
making it possible to introduce interaction of the singleton with $4d$ gravity
in the Cartan formalism. One can write a Lagrangian as a sum of the
singleton Lagrangians with the usual $4d$
matter Lagrangians of the SM, GR, and  further extensions,
\be
L=L^{Rac} + L^{Di} +L^{SM} +L^{GR} +\ldots\,.
\ee
As such, it is a sum of the three-form Lagrangians  $L^{Rac}$ and  $L^{Di}$
and four-form Lagrangians for the genuine $4d$ local fields contributing to $L^{SM}$ and
$L^{GR}$. Let us stress that since the Lagrangians are closed forms, the respective
actions are insensitive to the local variations of the three-cycles over which
$L^{Rac}$ and  $L^{Di}$ are integrated in the action,
\be
S=\int_{\Sigma_{Rac}^3} L^{Rac} + \int_{\Sigma_{Di}^3} L^{Di} +
\int_{M^4} ( L^{SM} +L^{GR} +\ldots)\,.
\ee

Because singletons are not local $4d$ fields, their direct scattering effects can unlikely be observable in the collider experiments. This implies that the presence
of singletons should not affect the local high-energy $4d$ SM physics, thus avoiding
a tension with the available experimental data. The same time, this
raises a question whether the singletons may yield observable phenomena whatsoever.

There is an alternative mechanisms that may play a role, however, namely
 the prominent Flato-Fronsdal theorem \cite{Flato:1978qz}
 stating that the tensor product of
two singleton fields (equivalently, their bilocal composits) amounts to the direct
sum of $4d$ massless fields including the massless spin two field, \ie graviton,
and a massless scalar in the singlet representation with respect to all inner symmetries,
\be
\label{FF}
S\bigotimes S = \sum_{s=0}^\infty \phi_{s,m=0}(x)= graviton +singlet\,\,scalar + \ldots\,.
\ee
For the extesion of Flato-Fronsdal theorem to the case with inner symmetries see
\cite{Konshtein:1988yg,Konstein:1989ij} while its harmonic analysis interpretation
was given in \cite{Iazeolla:2008ix}.
A related idea of bilocal fields was also put forward in \cite{deMelloKoch:2010wdf}.

That bilinears of singletons contain graviton and an inert massless scalar may have  observable consequences via, {\it e.g.,} an additionally induced gravitational field as well as invisible (dark) matter. In other words, though the $4d$ singleton fields cannot be localized themselves in some region of a galaxy, their nonlinear combinations can induce localizable fields. Of course,
to evaluate such effects one has to introduce and analyse interactions. The first step towards
gravitational interaction consists of the covariantization of $D_x$ in (\ref{dx}) beyond the flat
$AdS_{d+1}$ connection.
Note, however, that, as usual in the unfolded dynamics, this is only the first step
that demands further extension to respect the formal consistency in the sense of (\ref{cc}) beyond the linearized approximation. The simplest way towards solution to this problem is probably via  construction of the unfolded version of  $3d$ conformal gravity theory with
$3d$ massless matter. The $4d$ singleton theory then will result via the application of
the space-time metamorphoses mechanism of Section \ref{meta} to the
$3d$ conformal theory. For some progress in this direction within $3d$
conformal higher-spin gravity see \cite{Nilsson:2013tva,Nilsson:2015pua}, while unfolded formulation of conformal geometry was considered in \cite{Joung:2021bhf}.
The construction of the nonlinear $3d$ conformal higher-spin theory interacting with singletons was proposed in the recent papers \cite{Diaz:2024kpr,Diaz:2024iuz}.
The problem is to
evaluate the effect of condensation of the  gravitational and scalar fields in the galaxy induced by the singleton.

An intriguing group-theoretic singleton phenomenon found in
\cite{Nilsson:2018lof} is that the compactification of eleven-dimensional
supergravity on the seven sphere leads to singleton representations
along with the usual $4d$ relativistic representations in the spectrum.
Moreover, in  the squashed seven sphere case singletons were argued to be involved into
certain higgsing together with the usual $4d$ unitary representations. Since the field-theoretic interpretation of this phenomenon is still not clear it would be interesting to
apply the developed unfolded machinery to its further analysis. The formalism developed
in this paper may be most appropriate in this respect since
unfolded dynamics essentially maps the representation theory to dynamical equations.
Also a potentially useful  for  better understanding of this phenomenon
is the coset space approach to $AdS/CFT$ of  \cite{Ivanov:2003dt}. It would be
interesting to elaborate  more on its relation to the unfolded dynamics approach used in this paper.

Note that Flato-Fronsdal theorem suggests that the tensor product of singletons contains massless fields of all spins. This is a group-theoretic fact true at the free field level
that respects higher-spin symmetries. In the situation considered in this paper
with singletons interacting with the fields of SM and gravity this is no-longer
true and the tail of higher-spin fields is anticipated  to be deformed to some kind of effective
interactions. On the other hand, since lower-spin relativistic and inner symmetries remain unbroken, the graviton and scalar field contributions to (\ref{FF}) are anticipated to survive.

It is important to stress that, to be dynamically active, singleton should live
in the $(A)dS$  space. In other words the proposed construction
 can  only be working in presence of {dark energy}. More precisely, it works
 directly in the $sp(4)\sim o(3,2)$ invariant $AdS_4$ space with the
 negative cosmological constant $\Lambda$ while dark energy  associated with the $dS_4$ space has opposite sign of $\Lambda$.  In $dS_4$, the singleton Lagrangians contain imaginary unit via
 $\lambda := \sqrt{-\Lambda}$. However, as discussed in Section \ref{ext}, by an appropriate doubling  of fields one can make the model Hermitean. The presence of
 exponential solutions is natural in the $dS_4$  regime of
 expanding Universe. (Note  that singleton   in the flat space was argued to be a constant \cite{Flato:1986uh,Ponomarev:2021xdq,Bekaert:2022oeh}.)

 The presence of complex coefficients in the singleton equations
  may be related to such a persisting problem in cosmology and
 high-energy physics as baryon asymmetry. Indeed, these properties
  fit the prominent Sakharov necessary conditions \cite{Sakharov:1967dj}  for baryon asymmetry (for  review see, {\it e.g.}, \cite{Rubakov:1996vz}).
Positive cosmological constant may provide a non-equilibrium regime. Moreover, singletons endowed with appropriate inner structure may induce violation of the baryon number  conservation.
 The presence of complex coefficients in the Lagrangian may induce $CP$ violation.
Though, because of its $4d$ nonlocality,  the singleton field can hardly be seen
in usual scattering phenomena, it can affect the formation of a charged matter
via a nonlinear manifestation of the Flato-Fronsdal theorem.

A related comment is that being non-localisable in the $4d$ space
such fields
behave as background charges in the effective field theory if they have some non-zero VEVs responsible for violation of some global symmetries beyond the list of those manifest in the SM. In that case the proposed idea may have some relation to another Dirac's idea of evolution of fundamental (cosmological) constants \cite{Dirac:1937ti} via  evolution of the singleton fields.

\section{Conclusion}
\label{conc}

The goal of this paper is to point out that there is an unusual type of $4d$ relativistic
field, the Dirac singleton, that exists in presence of cosmological constant (dark energy). The new result of the paper is the formulation of its dynamics directly in the $4d$ space-time in a way free of the necessity of factorisation of the bulk modes as in the dipole approach of \cite{Flato:1986uh}. The singleton matter is unusual in the sense that it is $4d$ non-localisable. This is closely related to the fact that singleton forms an infinite-dimensional representation of the $4d$
Lorentz group, that does not mean, however, that singleton has more degrees of freedom than usual relativistic fields associated with finite-dimensional tensor-spinor Lorentz representations. Just other way around, singleton is essentially a $3d$ conformal field of usual type.

In this respect it is interesting to compare the proposed construction  with an alternative recent
dark matter candidate suggested by Bogomolny \cite{Bogomolny:2024ger} based on another nonstandard relativistic matter proposed by Dirac \cite{Dirac:1971cy}.
The difference  is that the latter equation describes infinitely many degrees
of freedom in four dimension while singleton, being a $3d$ field, carries less
than one usual relativistic field. It should be noted that
Flato and Fronsdal (see \cite{Flato:1999yp} and references therein) suggested
to use singleton as a kind of constituent matter with  usual relativistic fields realized as its composites. In this paper singleton is described directly in four dimensions as a relativistic field coexisting with other relativistic fields of usual types.

Note that one can treat analogously other conformal fields in
appropriate space-time dimensions. For instance, $4d$ spin-one massless field is conformal and hence can be uplifted to a relativistic singleton-type field in $(A)dS_5$ as discussed in \cite{Ferrara:1997dh}. More general types of singletons
associated with unitary conformal fields in even dimensions can also be considered
\cite{Bekaert:2009fg,Bekaert:2011js}.

Let us stress again that,
since the singleton field is nonlocal from the $4d$ perspective,
it can hardly be observed via a radiation process but can affect the
surrounding  gravitational field to be observed via the matter motion,
that may induce a dark matter-like effect. Another potential application
discussed in Section \ref{BSM} is for the baryon asymmetry explanation.
More generally, singletons may admit some cosmological manifestations
affecting the cosmological evolution as well. This is  what the scientific interests of Alexei Starobinsky were focused on.

\section*{Acknowledgments}
I wish to thank Konstantin Alkalaev, Mikhail Alfimov, Andrei Barvinsky, Olga Gelfond, Sergey Godunov, Carlo Iazeolla, Evgeny Ivanov, Andrei Mironov, Nikita Misuna, Roman Nevzorov, Bengt Nilsson, Dmitry Ponomarev, Aleksandr Pukhov,  Arkady Tseytlin,  Boris Voronov, and Mikhail Vysotsky for  useful comments.
I am grateful for hospitality to Ofer Aharony,
Theoretical High Energy Physics Group of Weizmann Institute of Science where a
part of this work has been done. This work was supported by Theoretical Physics and
Mathematics Advancement Foundation BASIS Grant No 24-1-1-9-5.


\begin{thebibliography}{10}

\bibitem{Dirac:1963ta}
P.~A.~M.~Dirac,
J. Math. Phys. \textbf{4} (1963), 901-909.

\bibitem{Flato:1986uh}
M.~Flato and C.~Fronsdal,
Commun. Math. Phys. \textbf{108} (1987), 469.

\bibitem{Sahni:1999gb}
V.~Sahni and A.~A.~Starobinsky,
Int. J. Mod. Phys. D \textbf{9} (2000), 373-444
[arXiv:astro-ph/9904398 [astro-ph]].

\bibitem{Ehrman}
 J.~B.~Ehrman,
Proc. Cambridge Philos. Soc., 53, 290 (1957).

\bibitem{Flato:1978qz}
M.~Flato and C.~Fronsdal,
Lett. Math. Phys. \textbf{2} (1978), 421-426.

\bibitem{Starinets:1998dt}
A.~Starinets,
Lett. Math. Phys. \textbf{50} (1999), 283-300
[arXiv:math-ph/9809014 [math-ph]].

\bibitem{Flato:1999yp}
M.~Flato, C.~Fronsdal and D.~Sternheimer,
[arXiv:hep-th/9901043 [hep-th]].


\bibitem{Sezgin:1983ik}
E.~Sezgin,
Phys. Lett. B \textbf{138} (1984), 57-62


\bibitem{Blencowe:1987bn}
M.~P.~Blencowe and M.~J.~Duff,
Phys. Lett. B \textbf{203} (1988), 229-236.

\bibitem{Nicolai:1988ek}
H.~Nicolai, E.~Sezgin and Y.~Tanii,
Nucl. Phys. B \textbf{305} (1988), 483-496.

\bibitem{Bergshoeff:1988jm}
E.~Bergshoeff, A.~Salam, E.~Sezgin and Y.~Tanii,
Phys. Lett. B \textbf{205} (1988), 237-244.

\bibitem{Bergshoeff:1989av}
E.~Bergshoeff, E.~Sezgin and Y.~Tanii,
Int. J. Mod. Phys. A \textbf{5} (1990), 3599-3616.

\bibitem{Bergshoeff:1988jx}
E.~Bergshoeff, A.~Salam, E.~Sezgin and Y.~Tanii,
Nucl. Phys. B \textbf{305} (1988), 497-515.

\bibitem{Nilsson:2018lof}
B.~E.~W.~Nilsson, A.~Padellaro and C.~N.~Pope,
JHEP \textbf{07} (2019), 124
[arXiv:1811.06228 [hep-th]].

\bibitem{Duff:2025tot}
M.~J.~Duff, B.~E.~W.~Nilsson and C.~N.~Pope,
[arXiv:2502.07710 [hep-th]].

\bibitem{Maldacena:1997re}
J.~M.~Maldacena,
  Adv.\ Theor.\ Math.\ Phys.\  {\bf 2} (1998) 231
  [Int.\ J.\ Theor.\ Phys.\  {\bf 38} (1999) 1113]
  [arXiv:hep-th/9711200].

\bibitem{Gubser:1998bc}
S.~S.~Gubser, I.~R.~Klebanov and A.~M.~Polyakov,
  Phys.\ Lett.\  B {\bf 428}, 105 (1998)
  [arXiv:hep-th/9802109].

\bibitem{Witten:1998qj}
E.~Witten,
  Adv.\ Theor.\ Math.\ Phys.\  {\bf 2}, 253 (1998)
  [arXiv:hep-th/9802150].


\bibitem{Sezgin:2002rt}
E.~Sezgin and P.~Sundell,
Nucl. Phys. B \textbf{644} (2002), 303-370
[erratum: Nucl. Phys. B \textbf{660} (2003), 403-403]
[arXiv:hep-th/0205131 [hep-th]].

\bibitem{KP}I.~R.~Klebanov and A.~M.~Polyakov, {\it Phys.\ Lett.\ }
{\bf B550} (2002) 213, {\tt hep-th/0210114}.

\bibitem{Vasiliev:2014vwa}
M.~A.~Vasiliev,
Lect. Notes Phys. \textbf{892} (2015), 227-264
[arXiv:1404.1948 [hep-th]].

\bibitem{Bellucci:2002ji}
S.~Bellucci, E.~Ivanov and S.~Krivonos,
Phys. Rev. D \textbf{66} (2002), 086001
[erratum: Phys. Rev. D \textbf{67} (2003), 049901]
[arXiv:hep-th/0206126 [hep-th]].

\bibitem{Ivanov:2003dt}
E.~Ivanov,
Theor. Math. Phys. \textbf{139} (2004), 513-528
[arXiv:hep-th/0305255 [hep-th]].


\bibitem{Flato:1986bf}
M.~Flato and C.~Fronsdal,
Phys. Lett. B \textbf{172} (1986), 412-416.

\bibitem{Flato:1988pz}
M.~Flato and C.~Fronsdal,
J. Geom. Phys. \textbf{5} (1988), 37-61.

\bibitem{Flato:1990yt}
M.~Flato and C.~Fronsdal,
J. Math. Phys. \textbf{32} (1991), 524-531.

\bibitem{Flato:1998iy}
M.~Flato and C.~Fronsdal,
Lett. Math. Phys. \textbf{44} (1998), 249-259.

\bibitem{Vasiliev:1988xc}
M.~A.~Vasiliev,
Phys. Lett. B \textbf{209} (1988), 491-497.

\bibitem{Vasiliev:2005zu}
M.~A.~Vasiliev,
Int. J. Geom. Meth. Mod. Phys. \textbf{3} (2006), 37-80
[arXiv:hep-th/0504090 [hep-th]].

\bibitem{Vasiliev:2001zy}
M.~A.~Vasiliev,
Phys. Rev. D \textbf{66} (2002), 066006
[arXiv:hep-th/0106149 [hep-th]].

\bibitem{Misuna:2024dlx}
N.~Misuna,
[arXiv:2408.13212 [hep-th]].


\bibitem{lag} C.~Fronsdal, Massless Particles, Ortosymplectic Symmetry and Another Type
of Kaluza-Klein Theory, Preprint UCLA/85/TEP/10, in Essays on Supersymmetry, Reidel, 1986 (Mathematical Physics Studies, v.8).

\bibitem{Bandos:1999qf}
I.~A.~Bandos, J.~Lukierski and D.~P.~Sorokin,
Phys. Rev. D \textbf{61} (2000), 045002
[arXiv:hep-th/9904109 [hep-th]].

\bibitem{Bandos:2005mb}
I.~Bandos, X.~Bekaert, J.~A.~de Azcarraga, D.~Sorokin and M.~Tsulaia,
JHEP \textbf{05} (2005), 031.
[arXiv:hep-th/0501113 [hep-th]].

\bibitem{Kogan:1999bn}
I.~I.~Kogan,
Phys. Lett. B \textbf{458} (1999), 66-72
[arXiv:hep-th/9903162 [hep-th]].

\bibitem{Ohl:2012bk}
T.~Ohl and C.~F.~Uhlemann,
JHEP \textbf{05} (2012), 161
[arXiv:1204.2054 [hep-th]].

\bibitem{Shaynkman:2000ts}
O.~V.~Shaynkman and M.~A.~Vasiliev,
Theor. Math. Phys. \textbf{123} (2000), 683-700
[arXiv:hep-th/0003123 [hep-th]].


\bibitem{Aharony:2010ay}
O.~Aharony, D.~Marolf and M.~Rangamani,
JHEP \textbf{02} (2011), 041
[arXiv:1011.6144 [hep-th]].

\bibitem{Nilsson:2012ky}
B.~E.~W.~Nilsson,
[arXiv:1203.5090 [hep-th]].

\bibitem{Shaynkman:2004vu}
O.~V.~Shaynkman, I.~Y.~Tipunin and M.~A.~Vasiliev,
Rev. Math. Phys. \textbf{18} (2006), 823-886
[arXiv:hep-th/0401086 [hep-th]].


\bibitem{Shaynkman:2001ip}
O.~V.~Shaynkman and M.~A.~Vasiliev,
Theor. Math. Phys. \textbf{128} (2001), 1155-1168
[arXiv:hep-th/0103208 [hep-th]].


\bibitem{Vasiliev:2012vf}
M.~A.~Vasiliev,
J. Phys. A \textbf{46} (2013), 214013
[arXiv:1203.5554 [hep-th]].

\bibitem{Misuna:2020fck}
N.~G.~Misuna,
JHEP \textbf{12} (2021), 172
[arXiv:2012.06570 [hep-th]].

\bibitem{Konshtein:1988yg}
S.~E.~Konshtein and M.~A.~Vasiliev,
Nucl. Phys. B \textbf{312} (1989), 402-418.

\bibitem{Konstein:1989ij}
S.~E.~Konstein and M.~A.~Vasiliev,
Nucl. Phys. B \textbf{331} (1990), 475-499.

\bibitem{Iazeolla:2008ix}
C.~Iazeolla and P.~Sundell,
JHEP \textbf{10} (2008), 022
[arXiv:0806.1942 [hep-th]].

\bibitem{deMelloKoch:2010wdf}
R.~de Mello Koch, A.~Jevicki, K.~Jin and J.~P.~Rodrigues,
Phys. Rev. D \textbf{83} (2011), 025006.

\bibitem{Nilsson:2013tva}
B.~E.~W.~Nilsson,
JHEP \textbf{09} (2015), 078
[arXiv:1312.5883 [hep-th]].

\bibitem{Nilsson:2015pua}
B.~E.~W.~Nilsson,
JHEP \textbf{08} (2016), 142
[arXiv:1506.03328 [hep-th]].

\bibitem{Joung:2021bhf}
E.~Joung, M.~g.~Kim and Y.~Kim,
JHEP \textbf{12} (2021), 092
[arXiv:2108.05535 [hep-th]].

\bibitem{Diaz:2024kpr}
F.~Diaz, C.~Iazeolla and P.~Sundell,
JHEP \textbf{09} (2024), 109
[arXiv:2403.02283 [hep-th]].

\bibitem{Diaz:2024iuz}
F.~Diaz, C.~Iazeolla and P.~Sundell,
JHEP \textbf{10} (2024), 066
[arXiv:2403.02301 [hep-th]].

\bibitem{Ponomarev:2021xdq}
D.~Ponomarev,
JHEP \textbf{06} (2021), 055
[arXiv:2104.02770 [hep-th]].

\bibitem{Bekaert:2022oeh}
X.~Bekaert, A.~Campoleoni and S.~Pekar,
Phys. Lett. B \textbf{838} (2023), 137734
[arXiv:2211.16498 [hep-th]].


\bibitem{Sakharov:1967dj}
A.~D.~Sakharov,
Pisma Zh. Eksp. Teor. Fiz. \textbf{5} (1967), 32-35.

\bibitem{Rubakov:1996vz}
V.~A.~Rubakov and M.~E.~Shaposhnikov,
Usp. Fiz. Nauk \textbf{166} (1996), 493-537
[arXiv:hep-ph/9603208 [hep-ph]].


\bibitem{Dirac:1937ti}
P.~A.~M.~Dirac,
Nature \textbf{139} (1937), 323.

\bibitem{Bogomolny:2024ger}
E.~Bogomolny,
Universe \textbf{10} (2024) no.5, 222
[arXiv:2406.01654 [hep-th]].

\bibitem{Dirac:1971cy}
P.~A.~M.~Dirac,
Proc. Roy. Soc. Lond. A \textbf{322} (1971), 435-445.

\bibitem{Ferrara:1997dh}
S.~Ferrara and C.~Fronsdal,
Class. Quant. Grav. \textbf{15} (1998), 2153-2164
[arXiv:hep-th/9712239 [hep-th]].

\bibitem{Bekaert:2009fg}
X.~Bekaert and M.~Grigoriev,
SIGMA \textbf{6} (2010), 038
[arXiv:0907.3195 [hep-th]].

\bibitem{Bekaert:2011js}
X.~Bekaert,
[arXiv:1111.4554 [math-ph]].


\end{thebibliography}
\end{document}